\documentclass{ws-procs975x65}

\begin{document}

\title{A SUMMARY: QUANTUM SINGULARITIES IN STATIC AND CONFORMALLY STATIC SPACETIMES}

\author{DEBORAH A. KONKOWSKI}

\address{Department of Mathematics, U.S. Naval Academy\\
Annapolis, Maryland, 21012, USA\\
E-mail: dak@usna.edu}

\author{THOMAS M. HELLIWELL}

\address{Department of Physics, Harvey Mudd College\\
Claremont, California, 91711, USA\\
E-mail: helliwell@HMC.edu}

\begin{abstract}
This is a summary of how the definition of quantum singularity is extended from static
space-times to conformally static space-times. Examples are given.
\end{abstract}

\bodymatter

\section{Introduction}
The question addressed in this review is: What happens if instead of classical particle paths (time-like and null geodesics) one uses quantum mechanical particles to identify singularities in conformally static as well as static spacetimes? A summary of the answer is given together with a couple of example applications. This conference proceeding is based primarily on an article by the authors \cite{HK1}.

\section{Types of Singularities}

\subsection{Classical Singularities}
A classical singularity is indicated by incomplete geodesics or incomplete paths of bounded acceleration \cite{HE, Geroch} in a maximal spacetime. Since, by definition, a spacetime is smooth, all irregular points (singularities) have been excised; a singular point is a boundary point of the spacetime. There are three different types of singularity \cite{ES}: quasi-regular, non-scalar curvature and scalar curvature. Whereas quasi-regular singularities are topological, curvature singularities are indicated by diverging components of the Riemann tensor when it is evaluated in a parallel-propagated orthonormal frame carried along a causal curve ending at the singularity.

\subsection{Quantum Singularities}
A spacetime is QM (quantum-mechanically) nonsingular if the evolution of a test scalar wave packet, representing the quantum particle, is uniquely determined by the initial wave packet, manifold and metric, without having to put boundary conditions at the singularity\cite{HM}. Technically, a static ST (spacetime) is QM-singular if the spatial portion of the Klein-Gordon operator is not essentially self-adjoint on $C_{0}^{\infty}(\Sigma)$ in $L^2(\Sigma)$ where $\Sigma$ is a spatial slice. This is tested (see, e.g., Konkowski and Helliwell \cite{HK1}) using Weyl's limit point - limit circle criterion \cite{RS, Weyl} that involves looking at an effective potential asymptotically at the location of the singularity. Here a limit-circle potential is quantum mechanically singular, while a limit-point potential is quantum mechanically non-singular. 

\par This definition of quantum singularity has been utilized in the analysis of several timelike spacetime singularities; three examples by the authors include asymptotically power-law spacetimes \cite{ HK3}, spacetimes with diverging higher-order curvature invariants \cite{HK1}, and a two-sphere singularity \cite{ HK5}.

\section{Conformally Static Space-Times}

\par  The Klein-Gordon with general coupling of a scalar field to the scalar curvature is given by

\begin{equation}
(\Box - \xi R)\Phi=M^2\Phi
\end{equation}

\noindent where $M$ is the mass if the scalar particle, $R$ is the scalar curvature, and $\xi$ is the coupling ($\xi=0$ for minimal coupling and $\xi=1/6$ for conformal coupling).
Using the natural symmetry of conformally static space-times the radial equation easily separates allowing it to be put into so-called Schr\"odinger form to identify the potential, allowing easy analysis of the quantum singularity structure.

\section{Friedmann-Robertson-Walker Space-Times with Cosmic String}

\par A metric modeling a Friedmann-Robertson-Walker cosmology with a cosmic string \cite{DS} can be written as

\begin{equation}
ds^2= a^2(t)( -dt^2 + dr^2 + \beta^2 r^2 d\phi^2 +dz^2)
\end{equation}

\noindent where $\beta=1-4\mu$ and $\mu$ is the mass per unit length of the cosmic string. This metric is conformally static (actually conformally flat).
Classically it has a scalar curvature singularity times when $a(t)$ is zero and a quasiregular singularity when $\beta^2\neq1$. We focus on resolving the timelike quasiregular singularity.

\par For the quantum analysis \cite{HK1}, the Klein-Gordon equation with general coupling can be separated into mode solutions \footnote{Only the time equation contains the coupling constant $\xi$.} with the radial equation changed to Schr\"odinger form,

\begin{equation}
u'' + (E - V(x))u = 0
\end{equation}

\noindent where $E$ is a constant, $x=r$, and the potential

\begin{equation}
V(x) = \frac{m^2 - \beta^2 /4}{\beta^2 x^2}
\end{equation}.

\noindent Near zero one can show that the potential $V(x)$ is limit point if $m^2/\beta^2 \geq 1$. So any modes with sufficiently large $m$ are limit point, but  $m=0$ is limit circle and thus generically this conformally static space-time is quantum mechanically singular.

\section{Roberts Solution}

\par The Roberts metric \cite{Roberts}

\begin{equation}
ds^2 = e^{2t}(-dt^2 + dr^2 + G^2(r) d\Omega^2)
\end{equation}

\noindent where $G^2(r) = 1/4[ 1+ p - (1 -p) e^{-2r}]( e^{2r} - 1)$ is conformally static, self-similar, and spherically symmetric.
It has a classical scalar curvature singularity at $r = 0$ for $0 < p < 1$ that is timelike.

The massive minimally coupled Klein-Gordon equation can be separated into mode solutions and by changing both dependent and independent variables ($r=x$), we get an appropriate inner product and a one-dimensional Schr\"odinger equation similar to Eq.(3) where again $E$ is a constant but, here, near zero, $V(x)$ goes like $-1/4x^2 < 3/4x^2$ so the potential is limit circle and there is a quantum singularity \cite{HK1}.

\section*{Acknowledgments}
One of us (DAK) thanks B. Yaptinchay for discussions.

\bibliographystyle{ws-procs975x65}
\bibliography{ws-pro-sample}

\begin{thebibliography}{9}
\bibitem{HK1} T.M. Helliwell and D.A. Konkowski, {\em Int. J. Mod. Phys. A} {\bf 26}, 3878 (2011).
\bibitem{HE} S.W. Hawking and G.F.R. Ellis, {\em The Large-Scale Structure of Spacetime} (Cambridge University Press, 1973).
\bibitem{Geroch}R. Geroch, {\em Ann. Phys.} {\bf 48}, 526 (1968) 
\bibitem{ES} G.F.R. Ellis and B.G. Schmidt, {\em Gen. Rel. Grav.} {\bf 8}, 915 (1977).
\bibitem{HM} G.T. Horowitz and D. Marolf, {\em Phys. Rev. D} {\bf 52}, 5670 (1995).
\bibitem{RS} M. Reed and B. Simon, {\em Functional Analysis} (Academic Press, 1972); M. Reed and B. Simon, {\em Fourier Analysis and Self-Adjointness} (Academic Press, 1972). 
\bibitem{Weyl} H. Weyl, {\em Math. Ann.} {\bf 68}, 220 (1910).
\bibitem{HK3} T.M. Helliwell and D.A. Konkowski, {\em Class. Quantum Grav.} {\bf 24}, 3377 (2007).
\bibitem{HK5} T.M. Helliwell and D.A. Konkowski, {\em Gen. Rel. Grav.} {\bf 43}, 695 (2011).
\bibitem{DS} P.C.W. Davies and V. Sahni, {\em Class. Quantum Grav.} {\bf 5}, 1 (1988).
\bibitem{Roberts} M.D. Roberts, {\em Gen. Rel. Grav.} {\bf 21}, 907 (1989).
\end{thebibliography}

\end{document}